\def\sles{\lower2pt\hbox{$\buildrel {\scriptstyle <}
   \over {\scriptstyle\sim}$}}
\def\sgreat{\lower2pt\hbox{$\buildrel {\scriptstyle >}
   \over {\scriptstyle\sim}$}}

\def\g{$\gamma$}
\def\G{\Gamma}
\def\etal{{\it etal. \ }}

\documentstyle[preprint,aps]{revtex}
\begin{document}

\title{$\gamma$-RAY BURSTS AND NEUTRON STAR
MERGERS\thanks{Supported by a BRF grant.}\thanks{To appear in the
Proceedings of the Lanczos Centenary: Editted by Moody Chu, Robert
Plemmons, David Brown, and Donald Ellison, SIAM publications.}} -
\author{Tsvi Piran}
\address{Racah Institute for Physics,
The Hebrew University,      Jerusalem, Israel, 91904}
\maketitle

\begin{abstract}
$\gamma$-ray bursts have baffled theorists ever since their accidental
discovery at the sixties.  We suggest that these bursts originate in
merger of neutron star binaries, taking place at cosmological
distances.  These mergers release $\approx 10^{54}ergs$, in what are
possibly the strongest explosions in the Universe.  If even a small
fraction of this energy is channeled to an electromagnetic signal it
will be detected as a grbs.  We examine the virtues and limitations of
this model and compare it with the recent Compton \g-ray observatory
results. We also discuss the potential application of grbs to study
cosmology and show that these burst might lead to a new and
independent determination of both $H_0$ and $\Omega$.
\end{abstract}

\section{Prologue: $\gamma$-Ray Bursts Circa 1973}

$\gamma$-ray bursts (grbs) were accidentally discovered ahead of their
time.  Had it not been for the need to verify the outer space treaty
of 1967 (which forbade nuclear experiments in space) we would not have
known about these bursts until well into the next century.  No one
would have proposed a satellite to look for such bursts, and had such
a proposal been made it would have surely turned down as too
speculative.  The VELA satellites with omni-directional detectors
sensitive to \g-ray pulses, which would have been emitted by a nuclear
explosion, were launched in the mid sixties to verify the outer space
treaty.  These satellites never detected any nuclear explosion.
However, as soon as the first satellite was launched it begun to
detect puzzling, perplexing and above all entirely unexpected bursts.
The lag between the arrival time of the pulses to different satellites
gave a directional information and  indicated  that the  sources are
outside the solar system.  Still, the bursts were kept secret for
several years, until Kelbsdal Strong and Olson described them in a
seminal paper\cite{Kle} in 1973.

\section{False Clues?}

The rapid fluctuation in the signal (less than 10ms) suggested a
compact source, a neutron star or a black hole.  Several other clues
focused the attention of theorists towards neutron stars at the disk
of the galaxy.

First, came an analytic estimate \cite{Sch} of the optical depth to
$\gamma \gamma \rightarrow e^+ e^-$.  For an impulsive source we have:
$\tau_{\gamma \gamma} \approx {\sigma_T F D^2 /R^2 m_e c^2}$ where F
is the fluence ($\approx 10^{-5} ergs/cm^2$ in the early detectors and
$\approx 10^{-7}ergs/cm^2$ in Compton-GRO), D is the distance to the
source and $R$ its size (expected to be less than $10^9$ from timing
arguments).  Since $\tau_{\gamma \gamma} > 1$ for $D > 100 pc$, it was
argued that the sources must be at the disk of the galaxy.  Otherwise,
it was argued an optical thick system will cool down and radiate its
energy in the x-ray uv or optical band and not as \g-rays.  The
non-thermal spectrum also indicated that the sources are optically
thin. Incidentally,, it was the confrontation between this argument
and the indications from Compton-GRO that grbs are cosmological (which
we discuss later) that have lead to claims that grbs require ``new
physics". We will see that it ain't necessarily so.

A very strong and long (1000 sec) burst was observed on March 5th
1979.  The position of the burst coincided with a SNR remnant in the
LMC supporting  the idea that grbs originates on neutron stars.

Another clue came from the observation of absorption lines
\cite{Maz,Yosh}.  The lines were interpreted as cyclotron lines in a
$10^{12}$G magnetic field, a field strength that is found only on
neutron stars.

These clues and others have led to the consensus that grbs
arise  on neutron stars in the disk of the galaxy\cite{HL}, possibly
in their magnetosphere.

\section{Bursts Distribution Circa 1991}

There were, however some indications that the sources might not be
galactic.  In 1975 Usov and Chibisov \cite{Uso} suggested to use a
logN-LogS test to check if the bursts have a cosmological origin.
Later, in 1983 van den Bergh \cite{van} analyzed the distribution of
the 46 bursts that were known at that time and from the isotropy of
this distribution he concluded that the sources are either local at
distances of less than half of the galactic disk scale height or
cosmological at redshift $z>0.1$ (See also \cite{Hb}).  The
cosmological solution was accepted with skepticism since with typical
fluencies of $10^{-5} ergs/cm^2$ the bursts require $10^{49} ergs$ if
they originate at distances larger than $100 Mpc$!  In 1986
Paczy\'nski \cite{Pac86} argued that the bursts are cosmological and
suggested that some of the burst are lensed by intervening galaxies
and that this will provide an observational test to the cosmological
hypothesis.  In 1989 Eichler, Livio, Piran and Schramm\cite{Ei} (see
also \cite{P90}) suggested that the bursts originate in neutron star
mergers at cosmological distances (The possibility that grbs might be
produced in neutron star mergers was also mentioned without specifying
a model by \cite{Bel,Pac86,Goo,GDN}).  However, in 1991, just before
the Compton-GRO results were announced, Atteia {\it etal.}
\cite{Atteia} reported (at a $ 3\sigma$ level) that the 244 bursts
observed by the spacecrafts Venera 13, 14 and Phebus  are concentrated
towards the galactic plane, suggesting a disk population after all.

\section{Neutron Star Binaries}

Another seemingly unrelated and unexpected discovery was make in 1975
by Hulse and Taylor \cite{Hul} who found a pulsar, PSR 1913+16, that
was orbiting around another neutron star.  No one has predicted that
such systems exist, but in retrospect it was not surprising.  More
than half of the stars are in binary systems. If some of these
binaries survive the two core collapses (and the supernovae
explosions) needed to produce the neutron stars they will end in a
binary pulsar.

The binary pulsar have proven to be an excellent laboratory for
testing General Relativity.  The binary system emits gravitational
radiation which is too week to be detected directly, but its
back reaction could be observed.  By carefully following the arrival
time of the pulsar's signals Taylor and his collaborator have measured
the pulsar's orbit. They have shown that  the binary spirals in just
in the right rate to compensate for the energy loss by gravitational
radiation emission (with two neutron stars, tidal interactions and
other energy losses  are negligible).  For PSR 1913+16 the spiraling
in takes place on a time scale of $\tau_{GR} = 3 \times 10^8$years, in
excellent agreement with the general relativistic prediction
\cite{Tay}.  These observations not only confirm the general
relativistic prediction, they  also assure us that the orbit of the
binary is indeed decreasing and that inevitably in $3 \times 10^8$
years  the two neutron stars will collide and merge!

\section{Source Count and Event Rate}

For many years only one binary pulsar was known. A simple estimate
based on the observation of one binary pulsar in several hundred
observed pulsars led Clark {\it etal.}  \cite{Clav} to conclude that
about 1 in 300 pulsars is in a  binary. With a pulsars' birth rate of
one in fifty years this led to a binary birth rate of one in $10^4$
years. Assuming a steady state, this is also the merger rate.  This
estimate ignored, however, selection effects in the detection of
binary pulsars vs.  regular ones. Specifically PSR1913+16 is an
extremely bright pulsar which is detectable from much larger distance
than an average  pulsar. Currently there are four known binary pulsars
and an analysis based on their  luminosities  and life times
\cite{NPS,Phi} suggests that  there are $\sim 10^4 ~-~10^5$ neutron
star binaries in the galaxy and that their merger rate is one per
$10^{6}$years per galaxy.  This corresponds to $\sim 100$ mergers per
year in galaxies out to a distance of 1 Gpc and about $10^3$ per year
to the horizon.  Narayan, Piran and Shemi \cite{NPS} also predict that
a similar or somewhat smaller population of neutron-star black hole
binaries will exist.

\section{Neutron Star Mergers}

It was immediately realized, after the discovery of PSR1913+16 that
the binary produces a  unique chirping gravitational radiation signal
during the last seconds  before the neutron stars merge.  These
signals are probably the best candidates for detection of
gravitational radiation.  However, these events are  rare and to
observe them (gravitationally) in our life time we must turn to
extragalactic events.  This is the aim of the advanced gravitational
radiation detectors like LIGO \cite{LIGO}.

As the strongest sources of gravitational radiation neutron star
mergers attracted the attention of relativists, but most astronomers
ignored them as being too rare to be of interest. Clark and Eardley
\cite{Clae} have shown that the binding energy released in a neutron
star binary merger is $\sim5 \times 10^{53} ~-~10^{54}ergs$, making
these events possibly the most powerful explosions in the Universe. A
significant fraction of this energy is emitted as gravitational
radiation, both prior and during the collision.  A very sophisticated
gravitational radiation detector, LIGO, is built to detect these
gravitational radiation signals. But it will be around the turn of the
century when it is operational.

As the neutron stars collide a shock forms and the stars heat up.
Most of the binding energy is emitted as neutrinos \cite{Clae}. The
neutrino burst is comparable or slightly stronger than a supernova
neutrino burst (such as the one detected by Kamiokande and IMB from
1987A). To detect extragalactic events at cosmological distances we
need a detector which is $\approx 10^8$ times larger than those
detectors. With regular supernova neutrino bursts being a hundred
times more frequent it is clear that these neutrino signals are not
the prime candidates for detection.

Neutron star mergers are hiding from us by emitting their energy in
two channels with extremely small cross sections.  If even a small
fraction of the energy is channeled to an electromagnetic signal, its
much large cross section will make it much easier to observe. For many
years, I kept wondering what are the possible observational
consequences of such events \cite{P90}.

\section{Energy Conversion}

Goodman, Dar and Nussinov \cite{GDN} suggested that the neutrino-anti
neutrino annihilation $\nu + \bar \nu \rightarrow e^+ + e^-$ converts
a small fraction of the neutrino supernova burst to electron-positron
pairs which in turn annihilate to \g-rays, heat the surrounding
envelop and provide the energy required to power the supernova shock
wave.  In 1989, Eichler, Livio, Piran and Schramm \cite{Ei} (see also
\cite{P90}) suggested that the same mechanism operates in neutron star
mergers and converts $\sim10^{-3}$ of the emitted energy to pairs and
\g-rays.  This corresponds to $10^{51} ergs$, roughly sufficient for
detection of the bursts from cosmological distances. Eichler {\it
etal.} \cite{Ei} used the old estimate of Clark {\it etal.}
\cite{Clav} for the merger rate and suggested that these events would
be detected by Compton observatory  as grbs.

More recently, alternative energy generation mechanism such as
magnetic field recombination \cite{NPP} or accretion onto the neutron
star \cite{Pac92a} have been proposed and it was argued that they
provide comparable amounts of energy.

\section{Fireballs and Relativistic Effects}
\subsection{Why Fireballs}

One of the most robust results in the theory of grbs is that any burst
which is at a cosmological distance must inevitably form a
``fireball''. If grbs are indeed cosmological they are initially
optically thick, as Schmidt \cite{Sch} have argued. How can there be a
\g-ray burst from such a source?  Goodman \cite{Goo} considered a
dense sphere of \g-ray photons and pairs, which he called a
``fireball".  He has shown that a cosmological grb source will quickly
form an extremely optically thick soup of electrons, positrons, and
photons, plus any baryons which may have been injected initially.
This optically thick relativistic fluid is referred to as a fireball.
The immense pressure in a fireball causes the fluid to expand
relativistically to a very large radius before the radiation can
finally escape.  The fireball will expand and cool, just like the
early Universe (unlike our Universe the gravitational force is
unimportant). As the fireball cools  its temperature drops with $T
\propto 1/R$ until the electron positions annihilate (the annihilation
is complete at $T \approx 20$  keV) and the radiation escapes. The
radiation fluid has reached in the meantime a relativistic velocity
relative to an observer at infinity and its Lorentz factor $\Gamma
\approx R_{esc}/R_0 \approx T_0/T_{esc} \approx 10^3 - 10^4$. The
escaping photons, which have a typical energy of 20 keV in the local
frame are blue shifted relative to an observer at infinity and their
observed energy is $\epsilon_{obs} \approx \G T_{esc} \approx T_0$, of
the same order as the initial energy. In this way the optical depth
argument which limited the distances to the sources is bypassed and
there is no need to introduce ``new physics" to explain grbs from
cosmological distances.

Paczy\'nski \cite{Pac86} have shown that similar effects take place if
the radiation is released in a quasi-stationary manner. In this case
the radiation flows out as a relativistic wind, with $T\propto 1/R$
and $\Gamma \propto R$. The radiation ceases to behave like a fluid
and escapes when $T \approx 20$ keV in the local frame. The escaping
x-ray photons are blueshifted to much higher energies in the observer
frame.

\subsection{Do Fireballs Work?}

The fireball model faces two serious objections: the origin of the
observed nonthermal spectrum and the effects of baryons.

There is no simple way to explain the non-thermal spectrum from a
fireball that passes an optically thick phase and termalizes.  It is
possible that different regions in a realistic, inhomogeneous fireball
move with significantly different Lorentz $\Gamma $ factors and that
the observed spectrum is a blending of thermal spectra to a non
thermal one. Simple calculations of the spectrum of a spherical
fireball \cite{SPN} show some deviation from a thermal spectrum, but
it is not large enough.  Alternatively one could hope that the
spectrum would become nonthermal in the transition from optically
thick to optically thin regimes.  However, this transition takes place
at $\approx 20$keV in the local frame. The energy injected from
annihilation at this stage is insignificant and the temperature is too
low for inverse Compton scattering to be effective.  It seems that
there is no clear mechanism that will modify the photons' black body
spectrum in this stage.

One expects that some baryons will be injected into the fireball.
Shemi and Piran \cite{SP} have shown that the baryons have two
effects.  For $10^{-11}M_\odot<M<10^{-8}M_\odot (E_0/10^{51}ergs)$ the
baryons dominate the opacity (long after all the pairs have
annihilated) without influencing the fireball's inertia. The fireball
continues to be optically thick until $\tau_g = \sigma_T M / R^2 = 1$.
This leads to a longer acceleration phase and to a larger final
Lorentz factor $\Gamma_f \approx R/R_0\approx T_0/T$.  However, the
final energy of the escaping radiation remains unchanged with
$\epsilon \approx \Gamma T \approx T_0$.

Larger baryonic load changes the dynamics of the fireball.  As the
fireball expands $\rho \propto R^{-3}$ while $e \propto r^{-4}$. If $M
> 10^{-8}M_\odot (E_0/10^{51}ergs)$ the baryonic rest mass will
dominate the energy density and the fireball's inertia before the
fireball becomes optically thin.  In these cases all the energy will
be used to accelerate the baryons with $E_K = Mc^2 \Gamma \approx (E_0
+ Mc^2) / (E_0 T / Mc^2 T_0 + 1)$.  The final outcome of a loaded
fireball will be relativistic expanding baryons with $\Gamma \approx
E_0 / M c^2$ and no radiation at all.

Several ideas have been proposed to avoid the baryonic load problem.
These include: (i) Separation of the radiation and the baryons due to
deviations from spherical symmetry - the radiation escaping along the
axis and the baryons being ejected preferably in the equatorial plane
\cite{PNS,Moch} and (ii) generation of a radiation fireball with very
small amounts of matter  via magnetic processes \cite{NPP}.

\subsection{Energy Conversion, Once More}

If the baryonic contamination is in the range $ 10^{-5} E_0<M c^2 <
0.1 E_0$ all the initial fireball energy will be converted to
extremely relativistic protons moving at a  Lorentz factors $10<
\Gamma \approx E_0 / M c^2< 10^5$.  M\'es\'zaros, and  Rees
\cite{MR1,MR2,MR3} (see also earlier work by Blandford and McKee
\cite{BM1,BM2} and more recent work by Katz \cite{Katz2,Katz3}
M\'es\'zaros, Laguna and Rees \cite{LMR}, Piran \cite{Pi93}
and Shemi \cite{Shem93}) suggested that
this energy could be converted back to \g-rays when this baryons
interact with the surrounding interstellar matter.  A shock, quite
similar to a SNR shock,  forms and it cools predominantly via
synchrotron emission in the x-rays.  The  x-ray photons will be
blueshifted to \g-rays in the observer frame due to the  relativistic
velocity of the fireball.  The relativistic motion will also lead to a
short time scale for the burst.  Alternatively, the accelerated
baryons could interact with a pre-merger wind that surrounds the
fireball \cite{NPP}.  In both cases the interaction with the
surrounding material will lead once more to the conversion of the
energy: from kinetic energy back to radiation.  Since this phase is
taking place in an optically thin region the  photons will  not
thermalize and the emerging spectra will be non thermal, as observed.
Thus, this process seems to resolve at one stroke both major
objections to the fireball scenario.

\section{Relativistic Bulk Motion - An Alternative?}

Several years ago Krolik  and Pier \cite{KroP} noticed that the large
optical depth problem, raised by Schmidt \cite{Sch} could be avoided
if the source is moving towards us at a relativistic velocity.
Several effects combine to remove this constraint. First the emission
is beamed with $\theta \approx 1 / \Gamma$ the anisotropy of the
emission  lowers the required density of photons at the source.  More
important is the fact that the observed photons  have been blue
shifted.  What is observed as \g-rays on earth is in fact x-rays or
even uv photons at the source.  At the source  only a minute fraction
of the photons is energetic enough  to produce pairs and the optical
depth problem to $\gamma\gamma \rightarrow e^+ e^-$ disappears. Krolik and
Pier have  obtained an estimate for the minimal bulk motion
required to explain the observed bursts which, depending on the location of
the bursts, leads to $\Gamma$ of hundreds or more.

A fireball is, in some sense, a variant of the Krolik and Pier idea.
Here we also observe blue shifted photons emitted in a rest frame that
is moving relativistically towards us.  However, the relativistic
velocity of the fireball is not due to a bulk motion of the source but
to relativistic expansion, which is an inevitable part of the fireball
scenario and follows from  the dynamics  of the model.  Since the
expansion is isotropic (at least in a spherical fireball) the radiation
is emitted isotropically. On the contrary, a moving source beams its
radiation in the direction of its bulk motion.

Kinematics attempts to explain grbs on the basis of bulk motion are
generally  misleading since they ignore the huge energy required to
produce a relativistic bulk motion of any macroscopic source. First it
is hard to reconcile the required high $\Gamma$ with the fact that the
highest observed relativistic motion in any other astronomical system
is less than ten and even this motions appears extremely rarely in
some AGNs' jets.  Moreover, The kinetic energy required, for example,
for a stellar mass source with a ``modest" $\Gamma$ of 100  is
$10^{56}$ergs.  It seems that any model that is based on an ad hoc
relativistic bulk motion of the source creates a severe problem - how
is the source accelerated which is as problematic as the phenomenon
that it attempts to explain.

\section{\g-ray Bursts Distribution Circa 1992}

The Compton \g-ray observatory was launched in the spring of 1991 (see
\cite{Paci} for a review).  It includes an omni-directional \g-ray
burst detector (BATSE) which, with a limiting sensitivity of $\approx
10^{-7} ergs/cm^2$, is the most sensitive detector of this kind flown.
By the summer of 1992  BATSE has detected more than 400 bursts, more
than all previous detectors combined.  BATSE is also capable of
obtaining a directional information on the bursts on its own.  Within
four month from its launch BATSE has collected enough data to conclude
that the distribution of grbs sources is isotropic \cite{Mee}.  The
average $V/V_{max}$ of the source is $\approx .33$ many $\sigma$ from
the value $0.5$ of a population distributed homogeneously in flat
space\cite{Mee}.  This show that the sources are not distributed
homogeneously in an Eucleadian space. They are either concentrated
towards us or alternatively they are distributed homogeneously in a curved
space-time and the observed inhomogeneity results from this curvature.

These  observations rule out all local galactic disk models.  A
possibility that was accepted by a small minority at first and gained
more and more support latter.  The observations are consistent with
three possible populations: (i) Cosmological population (ii) Galactic
halo population with a large core radius ($>50kpc$) and (iii) A
population, such as comets at the Oort cloud, centered around the
solar system.  We will turn to the second and third possibilities,
before summing up the status of the cosmological  model.

\subsection{Galactic Halo Models}

Galactic Halo models require a halo population with a large core
radius (to avoid an anisotropic enhancement towards the galactic
center). This is a new population of astronomical objects, which was
not seen elsewhere \cite{Pac92a}.  By now there have been several
suggestions how to form a neutron population of this kind.  These
include either ejection from the galactic disk  or formation in site.
However, the typical distances of a galactic  halo object, lead to
several difficulties which make such a location quite
unfavorable for production of grbs.

Approximately $10^{41} ergs$ are needed for bursts at the halo, quite
a large amount for a neutron star.  With a typical size of $10^6cm$
the optical thickness for $\gamma \gamma \rightarrow e^+ e^-$ is
$\approx 10^8$.  The energy requirement and the optical depth mean that
the low energy, optically thin neutron star models suggested for
galactic disk sources are inapplicable to grbs at the halo.  Furthermore,
galactic halo sources inevitably involve an opaque pair plasma
fireball, just like cosmological sources \cite{PS2}.  These fireballs
reach, however, lower relativistic Lorentz factors before becoming
optically thin or matter dominated. Those relatively moderate Lorentz
factors are unlikely to suffice for producing grbs.

\subsection{Local Population}

Typical objects in the solar system have a very small binding energy
per baryon and it is difficult to imagine a mechanism in which such
objects generate energies in the \g-ray range (see however
\cite{Katz}).  The only hope is probably via a magnetic phenomenon.
Solar flares do generate grbs which are detected by Compton-GRO (these
are identified by their location and spectrum \cite{Fish}). However,
comparison of the size and masses involved in these events make it
inconceivable that similar conditions can be achieved elsewhere in the
vicinity of the solar system, without leaving any other trace.

\subsection{Cosmological Population}

Several groups \cite{MP,P92,Der,Schm} have shown that a cosmological
population is compatible with the observed $V/V_{max}$ distribution.
The apparent concentration towards us is an artifact of a combination
of redshift effects and a possible cosmological evolution.
Depending of the
cosmological model and the source evolution we have $0.3<  Z_{av}  < 3
$.  For $\Omega=1$ and no evolution  $  Z_{av}   \approx 1$
\cite{P92}.

The cosmological model has a clear prediction \cite{P92,Pac92b,Pir92}:
a  positive correlation between the faintness of a burst (correlated
with distance) and redshift signatures through the burst duration and
spectrum.  This correlation could be masked by large
intrinsic variations among bursts, but should eventually be observed
when enough data accumulate.

The event rate needed to explain the observation is in an amazing
agreement with the rates estimated for neutron star mergers
\cite{NPS,Phi}.  Because of a historical  coincidence the forth binary
pulsar, PSR1534+12, which played a decisive role in the determination
of the merger rate \cite{NPS,Phi}, was discovered \cite{Wol} a few
month before Compton-GRO was launched and the prediction of the
neutron star merger rate were not influenced by the rates required to
explain the Compton-GRO results.

Several other cosmological  models were suggested after Compton-GRO
\cite{Car,McB,Hoy,Uso92}. Within the cosmological framework, the
neutron star merger scenario is the most conservative one possible.
It is the only one based on a source population that definitely
exists. We know its members will merge, we can be certain that huge
quantities of energy will be released in such mergers, and we find the
merger rate to be comparable to the observed burst rate.

\section{\g-ray Bursts Distribution Circa 1994}

BATSE on Compton-GRO has detected so far more than 700 grbs
\cite{Mee93}.  With more data and  better statistics the grb
distribution looks more isotropic than ever. $\langle V/V_{max}
\rangle $ converges to a value of $0.31$ demonstrating, if it was ever
necessary, that the preliminary value was not a statistical
fluctuation.  While only 5\% of the researcher present in the first
\g-ray bursts Huntsville conference in Oct 1991 supported the
cosmological hypothesis more than 50\% of those present in the second
meeting in Oct 1993 were in its favor. In two year the binary neutron
star merger model have gone the whole way from a crazy idea supported
by a few enthusiasts to become {\it the most conservative current
model} \cite{Rees}.

The improved data poses more and more problems to the Galactic Halo
model as it pushed the typical distance to a galactic halo source (to
be compatible with isotropy) farther and farther away. The current
data requires a core radius  $\sgreat 80$kpc \cite{Har2} which, as
predicted in 1992 \cite{Pir92}, is incompatible with the distribution
of the dark halo of the Galaxy (the latter being more concentrated
towards the galactic center).

Recently Norris \etal \cite{Nor} have found  the predicted
\cite{P92,Pac92b,Pir92} correlation  between the intensity and the
duration of the bursts and between the intensity and their hardness
ratio. This correlation is a clear indication of a cosmological
origin. It also provides an independent measure of the typical red
shift to a grb. The analysis suggests that the bursts originate from a
population with $Z_{max} \approx 1$, in agreement with analysis of the
distribution of intensity of the bursts!

It seems that the data support the cosmological hypothesis. However,
there are still claims that not everything is settled yet.  Quashnock
and Lamb \cite{QL,QLa} pointed out that there is a nearest neighbors
excess in the first BATSE catalogue, which they interpret as an
indication that grbs repeat. Clearly repetition of bursts from the
same location  rules out the neutron star merger model. The huge
energy budget required will make it quite difficult to construct any
cosmological repeating source (note that a  cosmological source
releasing several bursts  of $\approx10^{51}$ergs per year has an average
energy output of $\approx 10^{44}$ergs/sec which equals the luminosity
of a qso!).  In a different paper Quashnock and Lamb \cite{QL2,QLb} show
that if one divide the bursts to sub populations according to their
strength than one discovers that those sub-populations are
anisotropic. The combination of the two effects, suggests, according
to Quashnock and Lamb, that the bursts do originate at the galactic
disk.

Several problems with the repeater hypothesis caused, however, some
doubt.  Narayan and Piran \cite{NP,NP2} have shown that there is an
equal excess of furthers neighbors (that is bursts at the antipodal
location of other bursts). Such an excess cannot be explained by any
physical model and its existence suggests that both phenomena arise
from some inexplicable observational effect. Additionally, there is
some internal inconsistency between the narrow peak of the nearest
neighbors ($\approx 5^o$) and the typical positional errors of the
bursts which were much larger (the positional error depended on the
strength of the bursts and for weak bursts it could reach up to
$20^o$) (see also \cite{Har}). Preliminary studies of the full BATSE
catalog (containing more than 700 bursts) \cite{Mee93} show no nearest
or farthest neighbor excess and no anisotropy of sub populations.
Hence the repeater hypothesis and the sub population anisotropy are
probably ruled out (see however \cite{QLa,QLb}) and with this
disappears the last evidence in favor of a galactic disk origin.

\section{Clues Revisited}

Before concluding we turn once more to the  clues discussed earlier.
The optical depth problem disappeared in some sense and remained in
another.  Relativistic effects, due to the expansion of the fireball
\cite{Goo,Pac86}, were not taken into account in the original argument
\cite{Sch} which is flawed. The resulting spectrum from the expanding
fireball has the right energy range but to a first approximation it is
thermal.  It is a non-trivial (but not impossible) task to obtain a
nonthermal spectrum.  This problem is shared by all cosmological and
galactic halo models.

The March 5th event was one of three soft \g-ray repeaters, which have
a softer spectrum and produce  repeated bursts from the same source,
unlike all other sources \cite{HL}.  It is by now generally accepted
that these are most likely a different phenomenon.

The nature of the cyclotron lines has been fairly controversial since
they were first reported\cite{Lar,Hard}.  Mazets {\it etal.}
\cite{Maz} claimed that single ``cyclotron absorption lines" were
present in 20 bursts, with a broad distribution of line energies
(27--70 keV), but with only five lines having energies under 50 keV.
This is in conflict with the GINGA experiment which discovered three
systems of lines, all with nearly identical energies, all under 50 keV
\cite{Yosh}.  So far, no lines have been detected with any experiment
on the Compton Gamma Ray Observatory.

\section{Some Open Questions}

It would be misleading to draw a picture in which the grb
enigma has been completely solved. There are still some open
questions within the context of the cosmological model, within
the fireball model and within the more specific neutron star
merger model.

The first puzzle might also be the best clue to the problem.
Several groups \cite{Kou,GL,MNP} have shown that grbs can be
divided to two populations of short (shorter than $\approx 1$sec
vs. long bursts (longer than $\approx 1$sec) or equivalently
variable vs. smooth. The number count (LogN-LogS) distributions of
both sub populations agree with cosmological distributions.
More surprisingly the maximal peak luminosity of the bursts
in both sub population is equal to within a factor of two
even though the total energy released varies by more than
a factor of ten\cite{MNP}. This might be an accident but it
is more likely a possible clue for a mechanism
that controls the emission in the bursts, possible at the outer
edge of a relativistic fireball. At present there is no explanation
for that.

It is clear from the current understanding of fireballs that the
question of whether or not a fireball will produce a grb, and what
kind of a burst it produces, depends almost entirely on  the ratio $
E_0/M$.  This is because the asymptotic Lorentz factor of the baryons
is given by $\gamma\sim E_0/Mc^2 $.  Therefore, if $ E_0/M$ is too
small, i.e. if the baryonic load is too large, the flow will not reach
ultra relativistic velocities and there is unlikely to be a grb.  The
critical value seems to be $\Gamma \sgreat\ 10^2$, which produces a
rather strong limit on the amount of baryonic load
namely $M\ \sles\ 10^{-5} M_\odot (E_0/10^{51}$ergs).  It is
difficult to satisfy such a strict constraint.
This leads to an important set of
open questions namely: Even if $M$ is larger than the above limit, can
there still be isolated regions in a fireball where the local $E/M$
ratio is much higher than average and can such regions produce the
observed bursts? Can instabilities produce a phase separation between
low and high $E/M$  regions?  Can the same instabilities also explain
the extraordinary variety of burst profiles observed?

Within the neutron star merger model the baryonic load is a severe
problem.  However, the model also offers a simple solution. Mochkovich
\etal \cite{Moch} and Piran Narayan and Shemi \cite {PNS} pointed out
that due to the centrifugal force the matter forms a funnel along the
rotation axis in binary neutron star merger.  The baryonic load is
much lower within the funnel and this is a natural regime in which
$E/M$ will be high.  This suggestion has been confirmed by numerical
simulations of neutron star mergers that were carried our recently by
Davies \etal \cite{Dav}.

Davies \etal \cite{Dav} find that the coalescence, from
initial contact to the formation of an axially symmetric object, takes
only a few orbital periods. Some of the material from the two neutron
stars is shed, forming a thick disk around the central, coalesced
object. The mass of this disk depends on the initial neutron star
spins; higher spin rates resulting in greater mass loss, and thus more
massive disks. For spin rates that are most likely to be applicable to
real systems, the central coalesced object has a mass of $2.4M_\odot$,
which is tantalizingly close to the maximum mass allowed by any
neutron star equation of state for an object that is supported in part
by rotation.  Using a realistic nuclear equation of state Davies \etal
estimate the temperatures after the coalescence: the central object is
at a temperature of $\sim 10$MeV, whilst the disk is heated by shocks
to a temperature of 2-4MeV.  The disk is thick, almost toroidal; the
material having expanded on heating through shocks.  This disk
surrounds a central object that is somewhat flattened due to its rapid
rotation.

An almost empty centrifugal funnel forms around the
rotating axis and there is practically no material above the polar
caps.  This funnel provides a region in which a baryon free
radiation-electron-position plasma could form.  Neutrinos and
antineutrinos from the disk and form the polar caps would collide and
annihilate preferentially in the funnel (the energy in the c.m. frame
is larger when the colliding $\nu$ and $ \bar \nu$ approach at obtuse
angle, a condition that easily holds in the funnel). The numerical
computations do not show any baryons in the funnels. The resolution of
the computation is insufficient, however, to show that the baryonic
load in the funnel is as low as needed.  The neutrinos radiation
pressure on polar cap baryons can generate a baryonic wind that will
load the flow.  Estimates of this effect \cite{Dun,Woos} show that it
is negligible if the temperature on the polar caps is sufficiently
low. The estimated temperature from our computations is $\approx
2$MeV, which is marginal. Our temperature estimate is, however, least
certain in low temperature regions like this.  The current simulations
are clearly not accurate enough and do not include enough detailed
physics  to answer the question whether neutron star mergers could,
indeed, produce the required conditions for the initial fireball.

The numerical calculations support earlier suggestions \cite{PNS} that
the energy release in a binary neutron star merger is anisotropic, the
fireball appears as a jet along the rotation axis.  This poses an
immediate constraint on the model.  If the width of the jet is
$\theta$ than we observe grbs only from a fraction $2\theta^{-2}$ of
binary neutron star mergers.  The rates of grbs and binary neutron
star mergers agree only if $\theta \sgreat 0.2$ (unless the rate of
binary neutron star mergers is much higher than the current
estimates). A condition which at first glance is satisfied by the
funnel seen in the current numerical simulation.  The beaming will
also change of course the overall energy budget and lower the overall
energy by a factor of $\approx \theta^2$.

\section{\g-ray Bursts As Tools to Examine Cosmology}

If they bursts are cosmological then we can employ them to explore
cosmology, regardless of the nature of the sources.  Thus, grbs could
have much deeper and wider significance.  If, as some preliminary
tests indicate, grbs are indeed standard candles then they can be used
directly in a count tests.  Piran \cite{P92}  addressed this problem
for the first time in 1992, with a relatively poor set of data by
comparing the average $\langle V/V_{max}\rangle $ to the one resulting
from various theoretical cosmological distributions.  Later Wickaramasinghe
\etal \cite{Wik} and Mao \etal \cite{MNP} used the LogN-LogS test to obtain
a more precise measure. Currently this estimates have been done using
only the first BATSE catalogue.  The fuller catalog that contains
three times more bursts will naturally provide a better data set. In
principle the test is similar to previous attempts to measure $q_0$
(and hence $\Omega$) from galaxy counts or qso counts. In both cases
the test failed when it was discovered that the observed objects show
a significant density and luminosity evolution which screens the
cosmological effects.  Here, the LogN-LogS test can be improved
slightly by combining it with the correlation test of Norris \etal
\cite{Nor} that provide an {\it independent} test of $Z_{max}$ of the
observed grb population.  But history could repeat itself with
the grb population and evolution could mask cosmological effect here
as well \cite{P92}.

An additional improvement, which in principle could yield a
determination of Hubble's constant $H_0$, could take place towards the
end of the century if and when grbs will be found to coincide with
gravitational radiation emission from the merger. Shutz \cite{Schu}
pointed out that the gravitational radiation signal from a merger
provides a direct measure of the distance to the source. The
combination of the distance and redshift estimates would provide a new
and independent way to measure $H_0$.

\section{Epilogue: \g-ray Bursts Circa 2000}

At present there are no known optical counterparts to grbs.  Since
neutron star binaries might be ejected from dwarf galaxies,  we
predict \cite{NPP}, that grbs occur within a few tens of arcsecond
from dwarf galaxies and within but not necessarily at the center of
ellipticals. Optical identification of some parent galaxies, could
support this model and the location of the burst relative to the
galaxy could distinguish this model from other cosmological scenarios
that involve supermassive black holes or other objects located in the
centers of galaxies \cite{Car,McB,Hoy}.

The scenario makes one unique prediction: strong $\gamma$-ray bursts
should be accompanied by a gravitational wave signal
\cite{PNS,P92,NPP} (though the reverse need not necessarily be true if
the $\gamma$-rays are beamed).  These signals should be detected by
LIGO \cite{LIGO} when it becomes operational (hopefully by the year
2000).  A coincidence between gravitational radiation signals from the
final stages of the merger and grbs could prove or disprove this
model. It could also serve to increase the sensitivity of the
gravitational radiation detectors \cite{Koc}.  Hopefully, this
coincidence will be detected and the model will be confirmed when
gravitational radiation detectors will become operational at the turn
of the century. LIGO should provide good distance estimates to
individual bursts \cite{Schu} and should also pinpoint the exact time
of the merger. The distance measurement to the bursts could provide an
additional cosmological information and  in principle could lead to a
new and independent measurement of Hubble's constant $H_0$. ,

I would like to thank Ramesh Narayan for many helpful discussions.

\def\ApJ{{\it Ap. J.}}
\def\ApJL{{\it Ap. J. L.}}
\def\Nature{{\it Nature}}

\end{document}